\documentclass{article}
\usepackage{amssymb,amsmath,amsthm}
\newcommand{\ie}{\emph{i.e.}}
\newcommand{\eg}{\emph{e.g.}}
\newcommand{\cf}{\emph{cf}}
\newcommand{\Nat}{\mathbb{N}}
\newcommand{\Int}{\mathbb{Z}}

\newcommand{\Com}{\mathbb{C}}

\newcommand{\Dom}{D}
\newcommand{\Sobi}{W^{1,2}}
\newcommand{\Sobii}{W^{2,2}}

\newcommand{\Hilbert}{\mathcal{H}}
\newcommand{\PT}{\mathcal{PT}}
\newtheorem{Lemma}{Lemma}
\newtheorem{Proposition}{Proposition}
\newtheorem{Theorem}{Theorem}
\theoremstyle{remark}

\begin{document}
%
\title{\textbf{\Large
Closed formula for the metric in the Hilbert space
of a $\PT$-symmetric model
}}
\author{\textsc{
D.~Krej\v{c}i\v{r}\'{\i}k$^1$,
H.~B{\'\i}la$^{1,2}$
\ and \
M.~Znojil$^1$
}}
\date{
\footnotesize
\begin{quote}
\emph{
\begin{itemize}
\item[$1$]
Department of Theoretical Physics, Nuclear Physics Institute, \\
Academy of Sciences,
250\,68 \v{R}e\v{z} near Prague, Czech Republic
\smallskip
\item[$2$]
Faculty of Mathematics and Physics,
Charles University in Prague, \\
Ke Karlovu~3, 121\,16 Praha 2
\end{itemize}
}
\end{quote}
24 April 2006}
\maketitle
\begin{abstract}
\noindent
We introduce a very simple, exactly solvable
$\PT$-symmetric non-Hermit\-ian model with real spectrum,
and derive a closed formula for the metric operator
which relates the problem to a Hermitian one.
\end{abstract}
%
%
\section{Introduction}
%
In a way motivated by the needs of nuclear physics,
Scholtz, Geyer and Hahne~\cite{GHS}
established a general mathematical framework
for the consistent formulation of quantum mechanics
where a set of observables are represented
by bounded non-Hermitian operators $A_1,\dots,A_N$
with real spectra in a Hilbert space~$\Hilbert$.
In essence, they conjectured that in the similar situations
one may find a bounded positive Hermitian operator~$\Theta$,
called \emph{metric}, which fulfils
\begin{equation}\label{quasi}
  A_k^*\,\Theta = \Theta \,A_k
  \qquad\mbox{for all}\qquad
  k \in \{1,\dots,N\}
  \,,
\end{equation}
where~$A_k^*$ denotes the adjoint operator
of~$A_k$ in~$\Hilbert$.

Several years later, the notion of the metric operator~$\Theta$
re-emerged as a particularly useful mathematical tool
in the context of the so-called \emph{$\PT$-symmetric}
quantum mechanics \cite{Bender-Boettcher_1998,BBM}.
In this framework people usually paid attention
to the systems with a single observable,
\emph{viz}, with a Hamiltonian $A_1 \equiv H \neq H^*$
which possesses real spectrum
and for which the Schr\"odinger equation
is invariant under a simultaneous change of
spatial reflection~$\mathcal{P}$
and time reversal~$\mathcal{T}$.

In the current literature a lot of effort
has been devoted to the study of the particular models of $H$.
For their more detailed reviews and discussion
the reader is referred to the proceedings of the
\emph{International Workshops on Pseudo-Hermitian
Hamiltonians in Quantum Physics}
\cite{proceedings1,proceedings2,proceedings3}.
One finds that the construction of a non-trivial
operator $\Theta \neq I$, however difficult,
is a key to the correct probabilistic interpretation of
all the $\PT$-symmetric quantum systems
\cite{BBJ,Ali1,Ali2,Ali3}.
Indeed, it defines ``the physical" inner product
$
  (\cdot,\cdot)_\Theta:=(\cdot,\Theta\,\cdot)
$
which makes the Hamiltonian $H$ ``Hermitian" or,
in the language of~\cite{GHS}, \emph{quasi-Hermitian}.
For this reason, there
have been many attempts to calculate the metric operator~$\Theta$
for the various $\PT$-symmetric models of interest
\cite{Ali4,BMW,BBRR,BBJ1,BBJ2,BBJ2-erratum,Ali-Batal,BCMS,Ali5}.
Because of the complexity of the problem, however, it is not
surprising that most of the available formulae for~$\Theta$ are
just approximative, usually expressed as leading terms of
perturbation series \cite{Ali-Batal}.

The authors of~\cite{GHS} discussed
why our knowledge of the new inner product was necessary for the
evaluation of the physical predictions. They emphasized that the
theory endowed with it is a genuine quantum theory satisfying all
the necessary postulates. In a fairly recent continuation of this
discussion~\cite{Kretschmer-Szymanowski_2004} it has been
underlined that in the infinite-dimensional Hilbert spaces
$\Hilbert$ the requirement of the boundedness of the metric
operator~$\Theta$ plays a key role and that it deserves an
extremely careful analysis in the applications where a na{\"\i}ve
approach may lead to wrong results. In some sense, our present
paper may be read as a direct continuation of the rigorous
mathematical discussion in~\cite{Kretschmer-Szymanowski_2004}.

In particular we are going to illustrate here that our
understanding of~(\ref{quasi})
for unbounded operators~$H$ as the identity on functions
from the operator domain of~$H$ (\cf~(\ref{pseudo.rev}) below)
requires that~$\Theta$ maps the operator domain of~$H$ into the
operator domain of the adjoint~$H^*$. In such a setting we
imagined that the best way of finding a support for such an
argument can be sought in some exactly solvable $\PT$-symmetric
model. We decided to develop a new one -- such that its metric can
be obtained in a closed formula and in a rigorous manner.

The model we deal with in the present paper is one-dimensional,
defined in the Hilbert space
$$
  \Hilbert:=L^2\big((0,d)\big)
$$
where~$d$ is a given positive number.
In this Hilbert space we consider
the Hamiltonian~$H_\alpha$
which acts as the Laplacian, \ie,
$$
  H_\alpha\psi:=-\psi''
  \,,
$$
and satisfies the following Robin boundary conditions:
\begin{equation}\label{bc}
  \psi'(0) + i \alpha \;\! \psi(0) = 0
  \qquad\mbox{and}\qquad
  \psi'(d) + i \alpha \;\! \psi(d) = 0
  \,.
\end{equation}
Here~$\psi$ is a function from the Sobolev space $\Sobii((0,d))$
and~$\alpha$ is a real constant.
That is, the operator domain $\Dom(H_\alpha)$ consists
of functions with integrable (generalized) derivatives
up to the second order
and satisfying~(\ref{bc}) at the boundary points.
Because of the nature of the boundary conditions,
$H_\alpha$~is not Hermitian unless $\alpha=0$,
but it is $\PT$-symmetric with
the operators~$\mathcal{P}$ and~$\mathcal{T}$
being defined by $(\mathcal{P}\psi)(x):=\psi(d-x)$
and $\mathcal{T}\psi:=\overline{\psi}$, respectively.

It seems that our Hamiltonian~$H_\alpha$
represents the simplest $\PT$-symmetric model whatsoever.
The fact that the support of the non-Hermitian perturbation
is of measure zero invokes the $\PT$-symmetric models
\cite{AFK,Fei_2004,Znojil-Jakubsky_2001}
involving complex point interactions.
But our model is even simpler,
since it does not require any matching of solutions
known explicitly off the points
where the $\delta$-interaction is supported.

Indeed, the non-Hermiticity of~$H_\alpha$ enters
through the boundary conditions only,
while the Hamiltonian models a free quantum particle
in the interval $(0,d)$.
Consequently, the spectral problem for~$H_\alpha$
can be solved explicitly in terms of sines and cosines
(\cf~Section~\ref{Sec.spectrum} for more details).
Furthermore, an explicit form for the eigenfunctions
enables us to obtain a remarkably simple expression
for the metric operator:
\begin{Theorem}\label{Thm.metric}
Let~$\Theta(\alpha)$ be the linear operator
defined in~$\Hilbert$ by
\begin{equation}\label{Theta}
  \Theta(\alpha) := I
  + \phi_0^\alpha \, (\phi_0^\alpha,\cdot)
  + \Theta_0
  + i \alpha \;\! \Theta_1 + \alpha^2 \Theta_2
  \,,
\end{equation}
where~$I$ denotes the identity operator in~$\Hilbert$,
$(\cdot,\cdot)$ is the inner product on~$\Hilbert$,
antilinear in the first factor and linear in the second one,
\begin{equation}\label{phi0}
  \phi_0^\alpha(x) :=
  \sqrt{\frac{1}{d}} \, \exp{(i \alpha x)}
\end{equation}
and the operators~$\Theta_0$, $\Theta_1$ and~$\Theta_2$
acts in~$\Hilbert$ as
\begin{align}
  (\Theta_0\psi)(x)
  &:= - \frac{1}{d} \, (J\psi)(d)
  \,, \label{Theta0}
  \\
  (\Theta_1\psi)(x)
  &:= 2 \, (J\psi)(x) - \frac{x}{d} \, (J\psi)(d)
  - \frac{1}{d} \, (J^2\psi)(d)
  \,, \label{Theta1}
  \\
  (\Theta_2\psi)(x)
  &:= - (J^2\psi)(x) + \frac{x}{d} \, (J^2\psi)(d)
  \,, \label{Theta2}
\end{align}
with
\begin{equation}\label{J}
  (J\psi)(x) := \int_0^x \psi
  \,.
\end{equation}
Then~$\Theta(\alpha)$
is bounded, symmetric, non-negative
and satisfies
\begin{equation}\label{pseudo.rev}
  \forall \psi\in\Dom(H_\alpha), \qquad
  H_\alpha^* \;\! \Theta(\alpha) \, \psi
  = \Theta(\alpha) \;\! H_\alpha \;\! \psi
  \,.
\end{equation}
Furthermore, $\Theta(\alpha)$ is positive
if the condition
\begin{equation}\label{hypothesis}
  \alpha d/\pi \ \not\in \ \Int\!\setminus\!\{0\}
\end{equation}
holds true.
\end{Theorem}

Note that the metric $\Theta(\alpha)$ tends
to the identity operator~$I$ as $\alpha \to 0$,
which is expected due to the fact that~$H_0$
is nothing else than the (self-adjoint) Neumann Laplacian in~$\Hilbert$.
The condition~(\ref{hypothesis})
ensures that all the eigenvalues of~$H_\alpha$ are simple.
For simplicity, we do not consider
the degenerate cases in the present paper.

This paper is organized as follows.
In the following Section~\ref{Sec.Hamiltonian}
we introduce the Hamiltonian~$H_\alpha$
by means of its associated quadratic form;
this provides an elegant way of showing
that the operator is closed.
The spectral problem for~$H_\alpha$
is considered in Section~\ref{Sec.spectrum};
in particular, we show that the spectrum is real and discrete,
and write down the explicit eigenfunctions and eigenvalues.
Section~\ref{Sec.normalisation} contains the main idea
of the present paper;
namely, we observe that the eigenfunctions of~$H_\alpha$
are expressed in terms of Dirichlet and Neumann
complete orthonormal families in the interval $(0,d)$
and use a special normalization to simplify
the eigenfunctions of the adjoint~$H_\alpha^*$.
These enable us, in Section~\ref{Sec.metric},
to evaluate certain infinite series
defining the metric operator
and prove Theorem~\ref{Thm.metric}.
We conclude the paper by Section~\ref{Sec.end}
where we add several remarks
and discuss a possible extension of our model.

\section{The Hamiltonian}\label{Sec.Hamiltonian}
%
Let us first introduce the operator~$H_\alpha$ in a proper way.
We start with the associated sesquilinear form~$h_\alpha$
defined in the Hilbert space~$\Hilbert$
by the domain $\Dom(h_\alpha) := \Sobi((0,d))$
and by the prescription:
\begin{equation}\label{form}
  h_\alpha(\phi,\psi)
  := (\phi',\psi')
  + i \alpha \, \overline{\phi(d)} \, \psi(d)
  - i \alpha \, \overline{\phi(0)} \, \psi(0)
  \,.
\end{equation}
Here $(\cdot,\cdot)$ denotes the standard
inner product on~$\Hilbert$;
the corresponding norm will be denoted by $\|\cdot\|$.

Note that the boundary terms in~(\ref{form})
are well defined because
the domain of the quadratic form
is embedded in the space of uniformly continuous
functions on $(0,d)$
due to the Sobolev embedding theorem
\cite{Adams}.
It is also known that the Sobolev space $\Sobi((0,d))$
is dense in~$\Hilbert$; hence $h_\alpha$ is densely defined.
Moreover, the real part of~$h_\alpha$,
denoted by $\Re h_\alpha$,
is a densely defined, symmetric,
positive, closed sesquilinear form
(since it corresponds to the self-adjoint
Neumann Laplacian in~$\Hilbert$).
Of course, $h_\alpha$ itself is not symmetric unless $\alpha=0$,
however, it can be shown that it is sectorial and closed.
To see it, we use \cite[Thm.~VI.1.33]{Kato}
and prove that the imaginary part of~$h_\alpha$,
denoted by $\Im h_\alpha$,
is a small perturbation of $\Re h_\alpha$
in the following sense:
\begin{Lemma}\label{Lem.relative}
$\Im h_\alpha$ is relatively bounded with respect to $\Re h_\alpha$,
with
\begin{equation*}
  \big| (\Im h_\alpha)[\psi] \big|
  \leq
  \epsilon^{-1} \, \alpha^2 \, \|\psi\|^2
  + \epsilon \, (\Re h_\alpha)[\psi]
\end{equation*}
for all $\psi\in\Sobi((0,d))$
and any positive constant~$\epsilon$.
\end{Lemma}
\begin{proof}
Writing
$
  |\psi(d)|^2 - |\psi(0)|^2
  = \int_{0}^d \big(|\psi|^2\big)'
  = 2 \, \Re \big(\psi,\psi'\big)
$,
and applying the Schwarz and Cauchy inequalities
to the last term, we obtain the desired result.
\end{proof}

In view of the above properties
and the first representation theorem \cite[Thm.~VI.2.1]{Kato},
there exists a unique $m$-sectorial operator~$H_\alpha$
in~$\Hilbert$ such that
$
  h_\alpha(\phi,\psi) = (\phi,H_\alpha\psi)
$
for all $\phi\in\Dom(h_\alpha)$
and $\psi\in\Dom(H_\alpha)\subset\Dom(h_\alpha)$.
The operator domain $\Dom(H_\alpha)$ consists
of those functions $\psi \in \Dom(h_\alpha)$
for which there exists $\eta \in \Hilbert$ such that
$h_\alpha(\phi,\psi) = (\phi,\eta)$ holds
for every $\phi\in\Dom(h_\alpha)$.
Furthermore, using the ideas of \cite[Ex.~VI.2.16]{Kato},
it is possible to verify that indeed
\begin{equation}\label{operator}
\begin{aligned}
  H_\alpha\psi &= -\psi'' \,,
  \\
  \psi \in \Dom(H_\alpha)
  &= \big\{
  \psi \in \Sobii((0,d)) \ | \
  \psi \ \mbox{satisfies (\ref{bc})}
  \big\} \,.
\end{aligned}
\end{equation}
The above procedure also implies that
the adjoint operator~$H_\alpha^*$ is simply obtained
by the replacement $\alpha\mapsto-\alpha$.

Summing up the results, we obtain:
\begin{Proposition}\label{Prop.Hamiltonian}
The operator~$H_\alpha$ defined by~(\ref{operator})
is $m$-sectorial in~$\Hilbert$ and satisfies
$$
  H_\alpha^* = H_{-\alpha}
  \,.
$$
\end{Proposition}
%

\section{The spectrum}\label{Sec.spectrum}
%
An important property of an operator
being $m$-sectorial is that it is closed.
Then, in particular, the spectrum is well defined
by means of the resolvent operator.
We claim that our~$H_\alpha$ is an operator
with compact resolvent.
This can be seen by noticing that the Neumann Laplacian~$H_0$
(associated with $\Re h_\alpha$)
is an operator with compact resolvent
and by using the perturbation result of~\cite[Thm.~VI.3.4]{Kato}
together with Lemma~\ref{Lem.relative}.
Consequently, we know that the spectrum of~$H_\alpha$,
denoted by $\sigma(H_\alpha)$,
is purely discrete, \ie, it consists entirely
of isolated eigenvalues
with finite (algebraic) multiplicities.

The eigenvalue problem
$
  H_{\alpha}\psi = k^2 \psi
$,
with $k \in \Com$,
can be solved explicitly
in terms of sines and cosines.
In particular, the boundary conditions lead to
the following implicit equation for the eigenvalues:
\begin{equation}\label{implicit}
  (k^2-\alpha^2) \, \sin(k d) = 0
  \,.
\end{equation}
That is,
\begin{equation}\label{spectrum}
  \sigma(H_{\alpha})
  = \big\{\alpha^2\big\}
  \cup \big\{k_j^2\big\}_{j=1}^\infty
  \,,\qquad\mbox{where}\qquad
  k_j := j\pi/d \,.
\end{equation}
Hereafter we shall use the index $j\in\Nat$
to count the eigenvalues as in~(\ref{spectrum}),
with the convention that the eigenvalue for $j=0$
is given by~$\alpha^2$.

While the spectrum of~$H_{\alpha}$ is real,
it exhibits important differences
with respect to the spectra of self-adjoint
one-dimensional differential operators.
For instance, the spectrum of~$H_{\alpha}$ may not be simple
and even the lowest eigenvalue may be degenerate
for particular choices of~$\alpha$.
Notice also that~$H_{\alpha}$ coincides
with the spectrum of the Neumann Laplacian~$H_0$
up to the lowest (zero) eigenvalue
which is shifted to~$\alpha^2$.

In this paper we restrict to the non-degenerate case,
\ie, we make the hypothesis~(\ref{hypothesis}).
Then the eigenfunctions of~$H_\alpha$
corresponding to~(\ref{spectrum})
with the convention mentioned there
are given by
\begin{equation}\label{psi}
  \psi_j^\alpha(x) :=
  \begin{cases}
    A_0^\alpha \, \exp{(-i \alpha x)}
    & \mbox{if}\quad j=0 \,,
    \\
    A_j^\alpha \left(
    \cos(k_j x) - i \frac{\alpha}{k_j} \, \sin(k_j x)
    \right)
    &\mbox{if}\quad j \geq 1 \,,
  \end{cases}
\end{equation}
where each~$A_j$ is an arbitrary non-zero complex number.
In view of Proposition~\ref{Prop.Hamiltonian},
the spectrum of the adjoint~$H_\alpha^*$
coincides with~(\ref{spectrum})
and the corresponding eigenfunctions
are given by
\begin{equation}\label{phi}
  \phi_j^\alpha(x) :=
  \begin{cases}
    B_0^\alpha \, \exp{(i \alpha x)}
    & \mbox{if}\quad j=0 \,,
    \\
    B_j^\alpha \left(
    \cos(k_j x) + i \frac{\alpha}{k_j} \, \sin(k_j x)
    \right)
    &\mbox{if}\quad j \geq 1 \,,
  \end{cases}
\end{equation}
where each~$B_j$ is again an arbitrary non-zero complex number.

We collect the obtained spectral results
into the following proposition:
\begin{Proposition}\label{Prop.spectrum}
The spectrum of~$H_\alpha$ is real
and consists of discrete eigenvalues
specified in~(\ref{spectrum}).
If the condition~(\ref{hypothesis}) holds,
then all the eigenvalues have multiplicity one
and the corresponding eigenfunctions
are given by~(\ref{psi}).
\end{Proposition}
%

\section{Special normalization}\label{Sec.normalisation}
%
It follows directly by combining the eigenvalue problems
for~$H_\alpha$ and its adjoint that~$\phi_j^\alpha$
and~$\psi_k^\alpha$ are orthogonal to each other
provided $j\not=k$
and the non-degeneracy condition~(\ref{hypothesis}) holds.
The stronger result
\begin{equation}\label{orthonormality}
  \forall j,k\in\Nat, \quad
  (\phi_j^\alpha,\psi_k^\alpha) = \delta_{jk}
\end{equation}
will hold provided we normalize the eigenfunctions
in a special way.
Namely, (\ref{orthonormality})~follows by choosing
the coefficients~$A_j^\alpha$ and~$B_j^\alpha$
according to the equations
\begin{align}
  1 &=
  A_0^\alpha \, \overline{B_0^\alpha} \
  \frac{1-\exp(-2i\alpha d)}{2i\alpha}
  \,,
  \label{A0} \\
  1 &=
  A_j^\alpha \, \overline{B_j^\alpha} \
  \frac{(k_j^2-\alpha^2) \, d}{2 k_j^2}
  \qquad\mbox{for}\quad
  j \geq 1 \,.
  \label{Aj}
\end{align}
(If $\alpha=0$, the fraction in the first equation
should be understood as the expression
obtained after taking the limit $\alpha \to 0$.)
These equations can clearly be satisfied
as soon as~(\ref{hypothesis}) holds.

We still have a freedom in specifying~$A_j^\alpha$
and~$B_j^\alpha$. For further purposes, however,
we choose the coefficients~$B_j^\alpha$
in a very simple form by the requirements
\begin{equation}\label{normalisation}
  B_0 := \sqrt{1/d}
  \qquad\mbox{and}\qquad
  B_j := \sqrt{2/d}
  \qquad\mbox{for}\quad
  j \geq 1 \,,
\end{equation}
while we leave more complicated formula,
determined by the equations~(\ref{A0}) and~(\ref{Aj}),
for the coefficients~$A_j^\alpha$.
Then~$\phi_0^\alpha$ coincides with~(\ref{phi0})
and we have the decomposition
\begin{equation}\label{phi.bis}
  \phi_j^\alpha(x) =
  \chi_j^N(x) + i \frac{\alpha}{k_j} \, \chi_j^D(x)
  \qquad\mbox{for}\quad j \geq 1 \,,
\end{equation}
where $\{\chi_j^N\}_{j=0}^\infty$,
respectively $\{\chi_j^D\}_{j=1}^\infty$,
denotes the set of normalized eigenfunctions of the Neumann,
respectively Dirichlet, Laplacian in~$\Hilbert$:
\begin{equation*}
  \chi_j^N(x) :=
  \begin{cases}
    \sqrt{1/d}
    & \mbox{if}\quad j=0 \,,
    \\
    \sqrt{2/d} \, \cos(k_j x)
    & \mbox{if}\quad j \geq 1 \,,
  \end{cases}
  \qquad
  \chi_j^D(x) := \sqrt{2/d} \, \sin(k_j x)
  \,.
\end{equation*}
In addition to~(\ref{phi.bis}),
we also have the uniform convergence
$\phi_0^\alpha \to \chi_0^N$ as $\alpha \to 0$.
We point out the result we shall need later:
\begin{Proposition}\label{Prop.normalisation}
If the condition~(\ref{hypothesis}) holds true,
then the eigenfunctions $\psi_j^\alpha$ of~$H_\alpha$
and the eigenfunctions $\phi_j^\alpha$ of~$H_\alpha^*$
can be chosen in such a way that they satisfy
the biorthonormality relations~(\ref{orthonormality})
and the latter satisfy~(\ref{phi.bis}).
\end{Proposition}

The decomposition~(\ref{phi.bis}) plays a crucial role
in the subsequent section, mainly due to the fact
that $\{\chi_j^N\}_{j=0}^\infty$ and $\{\chi_j^D\}_{j=1}^\infty$
are well known to form \emph{complete orthonormal families}.
In particular, we have the expansions
\begin{equation}\label{expansion}
  \psi = \sum_{j=0}^\infty \chi_j^N \, (\chi_j^N,\psi)
  = \sum_{j=1}^\infty \chi_j^D \, (\chi_j^D,\psi)
\end{equation}
and the Parseval equalities
\begin{equation}\label{Parseval}
  \|\psi\|^2 = \sum_{j=0}^\infty |(\chi_j^N,\psi)|^2
  = \sum_{j=1}^\infty |(\chi_j^D,\psi)|^2
\end{equation}
for every $\psi\in\Hilbert$.

\section{The metric}\label{Sec.metric}
%
With an abuse of notation,
we initially define
\begin{equation}\label{metric}
  \Theta(\alpha) :=
  \sum_{j=0}^\infty \phi_j^\alpha \, (\phi_j^\alpha,\cdot)
\end{equation}
and show that this operator can be cast
into the form~(\ref{Theta}) with~(\ref{phi0})--(\ref{Theta2}).
In fact, using~(\ref{phi.bis}) and~(\ref{expansion}),
it is readily seen that~(\ref{Theta}) holds with
\begin{equation}\label{Theta0.bis}
  \Theta_0 := - \chi_0^N \, (\chi_0^N,\cdot)
\end{equation}
and
\begin{equation}\label{thetas}
  \Theta_1
  := \sum_{j=1}^\infty
  \frac{\chi_j^D \, (\chi_j^N,\cdot)-\chi_j^N \, (\chi_j^D,\cdot)}{k_j}
  \,, \qquad
  \Theta_2
  := \sum_{j=1}^\infty
  \frac{\chi_j^D \, (\chi_j^D,\cdot)}{k_j^2}
  \,.
\end{equation}
Recalling the definition~(\ref{J})
of the bounded integral operator~$J$ in~$\Hilbert$,
it is evident that the rank-one operator~(\ref{Theta0.bis})
can be expressed in terms of~$J$ as in~(\ref{Theta0}).
It remains to verify that~(\ref{thetas})
can be expressed as in~(\ref{Theta1}) and~(\ref{Theta2}).

First of all, we notice that the operator~(\ref{metric})
is well defined in the sense that~$\Theta_1$ and~$\Theta_2$
as defined in~(\ref{thetas})
are bounded linear operators in~$\Hilbert$.
This can be seen by using~(\ref{Parseval})
and the Schwarz inequality.
Actually, the series in~(\ref{thetas})
are uniformly convergent,
and~$\Theta_2$ can be written
as an integral Hilbert-Schmidt operator,
but we will not use these facts.
Our way to sum up the infinite series
is based on the following lemma:
\begin{Lemma}\label{Lem.sum}
$$
  \sum_{j=1}^\infty \frac{\chi_j^D(x) \, \chi_j^N(d)}{k_j}
  = - \frac{x}{d}
  \qquad\mbox{uniformly for all}\quad
  x\in[0,d]
  \,.
$$
\end{Lemma}
\begin{proof}
The series is uniformly convergent
due to Abel's uniform convergence test.
Let~$l$ denote the identity function on $(0,d)$,
\ie\ $l(x):=x$.
Using the expansion~(\ref{expansion})
and integrating by parts, we get
$$
  l = \sum_{j=1}^\infty \chi_j^D \, (\chi_j^D,l)
  = \sum_{j=1}^\infty \chi_j^D \,
  \big((-\chi_j^N/k_j)',l\big)
  = - \sum_{j=1}^\infty
  \chi_j^D \, \chi_j^N(d) \, d / k_j
  \,,
$$
where the last equality follows from the fact
that all $\chi_j^N$ with $j \geq 1$
are orthogonal to the constant function~$\chi_0^N$.
This concludes the proof.
\end{proof}

Since $J\psi$ is an indefinite integral of~$\psi$
and $(J\psi)(0)=0$, an integration by parts yields
for every $\psi\in\Hilbert$:
\begin{align*}
  (\chi_j^N,\psi)
  &= k_j \, (\chi_j^D,J\psi)
  + \chi_j^N(d) \, (J\psi)(d)
  \,,
  \\
  (\chi_j^D,\psi)
  &= - k_j \, (\chi_j^N,J\psi)
  = - k_j^2 \, (\chi_j^D,J^2\psi)
  - k_j \, \chi_j^N(d) \, (J^2\psi)(d)
  \,.
\end{align*}
Incorporating these identities into~(\ref{thetas})
and using~(\ref{expansion}) together with Lemma~\ref{Lem.sum},
we obtain the formulae~(\ref{Theta1}) and~(\ref{Theta2})
for~(\ref{thetas}).

Now we are in a position to prove Theorem~\ref{Thm.metric}.
\begin{proof}[Proof of Theorem~\ref{Thm.metric}]
The boundedness of~(\ref{Theta}) is clear;
in particular, crude estimates yield
$$
  \|\Theta(\alpha)\psi\|
  \leq (3 + 4 \alpha d + 2 \alpha^2 d^2) \, \|\psi\|
$$
for every $\psi\in\Hilbert$.

Integrating by parts, it is also easy
to check that the identity
\begin{equation}\label{symmetry}
  \big(\psi,\Theta(\alpha)\psi\big)
  = |(\phi_0^\alpha,\psi)|^2
  + \|\psi+i \alpha J\psi\|^2
  - |(\chi_0^N,\psi+i\alpha J\psi)|^2
\end{equation}
holds for every $\psi\in\Hilbert$,
where the right hand side is real-valued
and non-negative due to~(\ref{Parseval}).
This proves that~$\Theta(\alpha)$
is symmetric and non-negative.

Let us show that~$\Theta(\alpha)$ is positive
provided~(\ref{hypothesis}) holds.
If the right hand side of~(\ref{symmetry}) were equal to zero
with a non-zero $\psi\in\Hilbert$,
then the first Parseval equality in~(\ref{Parseval})
would imply that the function $\psi+i\alpha J\psi$ must be constant,
being orthogonal to all functions orthogonal to~$1$.
Consequently, $\psi$~is proportional to~$\psi_0^\alpha$
and an explicit calculation yields
$$
  |(\phi_0^\alpha,\psi)|
  = \left|\frac{\sin(\alpha d)}{\alpha d}\right|
  \, \|\psi\|
  \,,
$$
which is clearly positive for all~$\alpha$
satisfying~(\ref{hypothesis}).

Finally, let us comment on the identity~(\ref{pseudo.rev}).
Let $\psi\in\Dom(H_\alpha)$.
We first note that it straightforward to check that
$\Theta(\alpha)\psi$ belongs to $\Dom(H_\alpha^*)$,
so that the left hand side of~(\ref{pseudo.rev}) makes sense.
We also have
$$
  - (\Theta(\alpha) \;\! \psi)''
  = - \psi'' - 2i\alpha\;\!\psi' + \alpha^2 \psi
  + \alpha^2 \phi_0^\alpha (\phi_0^\alpha,\psi)
  = -\Theta(\alpha)\psi''
  \,.
$$
Here the first equality follows at once,
while the second one is not trivial,
but it can be verified
by using a number of integrations by parts.

This concludes the proof of Theorem~\ref{Thm.metric}.
\end{proof}
%

\section{Concluding remarks}\label{Sec.end}
%
\subsection{Alternative proofs of the reality of the spectrum}
Recall that $\PT$-symmetry itself
is not sufficient to guarantee the reality
of the spectrum of a non-Hermitian operator
(see, \eg, \cite{Znojil_2001,Znojil-Levai_2001}).
Moreover, the existing proofs of the reality
\cite{DDT,Langer-Tretter_2004,CGS,Shin}
are based on rather advanced techniques.
Therefore we find it interesting
that the reality of the eigenvalues of our Hamiltonian~$H_\alpha$
can be deduced directly from the structure of the operator,
without solving the eigenvalue problem explicitly.

To see it, we rewrite the eigenvalue problem
$
  H_{\alpha}\psi = k^2 \psi
$
using the unitary transform
$
  \psi \mapsto \phi_0^\alpha \psi := \phi
$
into the boundary value problem
\begin{equation}\label{Neumann}
\left\{
\begin{aligned}
  -\phi'' + 2 i \alpha \;\! \phi' + \alpha^2 \phi
  &= k^2 \phi
  &\qquad\mbox{in}&\qquad (0,d) \,,
  \\
  \phi'
  &= 0
  &\qquad\mbox{at}&\qquad 0,d \,.
\end{aligned}
\right.
\end{equation}
Now we multiply the first equation in~(\ref{Neumann})
by~$\overline{\phi''}$ and integrate over $(0,d)$.
We also multiply the complex conjugation
of the first equation in~(\ref{Neumann}) by~$\phi''$
and integrate over $(0,d)$.
Then we subtract the results and use various integrations
by parts together with the Neumann boundary conditions
to get the identity
\begin{equation*}
  \Im(k^2) \, \|\phi'\|^2 = 0
  \,.
\end{equation*}
Consequently, either the eigenvalue~$k^2$ is real
or the corresponding eigenfunction~$\phi$ is constant.
It remains to realize that also the latter
implies the former in view of~(\ref{Neumann}).

Finally, let us mention that~$H_\alpha$
can be reconsidered as a self-adjoint operator
in a Krein space~\cite{Langer-Tretter_2004}.
Then the reality of the spectrum of~$H_\alpha$
for $|\alpha|<\pi/d$
follows from \cite[Corol.~3.3]{Langer-Tretter_2004}.
An alternative proof of the reality of the spectrum
of~$H_\alpha$ for small~$\alpha$ also follows
from the perturbation result of~\cite{CGS}.

\subsection{Biorthonormal basis}
It is easily seen that the operator~$\Theta(\alpha)$
defined by~(\ref{metric})
formally satisfies~(\ref{pseudo.rev}),
with the inverse given by
%
$
  \Theta(\alpha)^{-1} =
  \sum_{j=0}^\infty \psi_j^\alpha \, (\psi_j^\alpha,\cdot)
  \,,
$
%
provided $\{\psi_j^\alpha\}_{j=0}^\infty$
and $\{\phi_j^\alpha\}_{j=0}^\infty$
fulfil in addition to~(\ref{orthonormality})
the following biorthonormal-basis-type relation:
\begin{equation}\label{basis}
  \forall \psi\in\Hilbert, \qquad
  \psi =
  \sum_{j=0}^\infty \psi_j^\alpha \, (\phi_j^\alpha,\psi)
  \,.
\end{equation}
By ``formally'' we mean that one has to justify
an interchange of summation and differentiation.
We did not pursue this direction in the present paper.
Instead, we summed up the infinite series~(\ref{metric})
using the special normalization~(\ref{normalisation})
leading to~(\ref{phi.bis}),
and checked that the resulting operator
indeed satisfies~(\ref{quasi})
in the sense of~(\ref{pseudo.rev}).

Nevertheless, let us show that
the expansion~(\ref{basis}) holds:
\begin{Proposition}
If the condition~(\ref{hypothesis}) holds true,
then the eigenfunctions $\psi_j^\alpha$ of~$H_\alpha$
and the eigenfunctions $\phi_j^\alpha$ of~$H_\alpha^*$
can be chosen in such a way that~(\ref{basis}) is satisfied.
\end{Proposition}
\begin{proof}
Assume the special normalization of Section~\ref{Sec.normalisation}.
Let us first verify that $\{\psi_j\}_{j=0}^\alpha$
is a basis of~$\Hilbert$,
\ie,
\begin{equation}\label{basis.bis}
  \forall \psi\in\Hilbert, \qquad
  \psi =
  \sum_{j=0}^\infty c_j^\psi \, \psi_j^\alpha
  \,,
\end{equation}
where $\{c_j^\psi\}_{j=0}^\infty$ is a unique
sequence of complex numbers.
Note that the equality in~(\ref{basis.bis})
should be understood as a limit in the norm
topology of~$\Hilbert$;
in particular, (\ref{basis.bis})~implies the weak convergence
\begin{equation}\label{weak}
  \forall\phi,\psi\in\Hilbert, \qquad
  (\phi,\psi) =
  \lim_{m\to\infty}
  \Big(\phi,\sum_{n=1}^m c_j^\psi \, \psi_j^\alpha\Big)
  \,.
\end{equation}
Substituting $\psi=0$ and $\phi=\phi_k^\alpha$, $k\in\Nat$,
into~(\ref{weak}), the biorthonormality relations~(\ref{orthonormality})
yield that (\ref{basis.bis})~with $\psi=0$ implies that all $c_j^0=0$.
At the same time,
$$
  \|\psi_j^\alpha-\chi_j^N\|^2
  = \alpha^2 \, \frac{k_j^2+\alpha^2}{(k_j^2-\alpha^2)^2}
  \qquad\mbox{for}\quad
  j \geq 1
$$
and since the right hand side behaves as
$\mathcal{O}(j^{-2})$ as $j\to\infty$,
we have
$$
  \sum_{j=0}^\infty \|\psi_j^\alpha-\chi_j^N\|^2
  < \infty
  \,.
$$
Consequently, $\{\psi_j\}_{j=0}^\alpha$
is a basis of~$\Hilbert$
due to \cite[Thm.~V.2.20]{Kato}.
Finally, substituting $\phi=\phi_k^\alpha$, $k\in\Nat$,
into~(\ref{weak}), the biorthonormality relations~(\ref{orthonormality})
yield that $c_j^\psi=(\phi_j^\alpha,\psi)$
for all $j\in\Nat$.
\end{proof}

The same argument also implies the following expansion:
\begin{equation*}
  \forall \psi\in\Hilbert, \qquad
  \psi =
  \sum_{j=0}^\infty \phi_j^\alpha \, (\psi_j^\alpha,\psi)
  \,.
\end{equation*}
\subsection{A more general model}
For simplicity, we required that~$\alpha$ was real
in the present paper.
A more general model is given by the following
more general $\PT$-symmetric boundary conditions:
\begin{equation}\label{bc.bis}
  \psi'(0) + (\beta+i\alpha) \;\! \psi(0) = 0
  \qquad\mbox{and}\qquad
  -\psi'(d) + (\beta-i\alpha) \;\! \psi(d) = 0
  \,,
\end{equation}
where~$\alpha$ and~$\beta$ are real constants.
A straightforward modification
of the approach of Section~\ref{Sec.Hamiltonian}
(\cf~also the first paragraph of Section~\ref{Sec.spectrum})
yields:
\begin{Proposition}
The operator~$H_{\alpha,\beta}$ defined in~$\Hilbert$ by
\begin{equation*}
\begin{aligned}
  H_{\alpha,\beta}\;\!\psi &= -\psi'' \,,
  \\
  \psi \in \Dom(H_{\alpha,\beta})
  &= \big\{
  \psi \in \Sobii((0,d)) \ | \
  \psi \ \mbox{satisfies (\ref{bc.bis})}
  \big\} \,,
\end{aligned}
\end{equation*}
is an $m$-sectorial operator with compact resolvent
and satisfies
$
  H_{\alpha,\beta}^* = H_{-\alpha,\beta}
$.
\end{Proposition}

The eigenvalue problem
$
  H_{\alpha,\beta}\;\!\psi = k^2 \psi
$,
with $k \in \Com$,
can again be solved in terms of sines and cosines,
and one gets the following implicit equation
for the eigenvalues:
\begin{equation*}
  \big[k^2-(\alpha^2+\beta^2)\big] \, \sin(k d)
  - 2 \, \beta \, k \, \cos(k d)
  = 0
  \,.
\end{equation*}
The main difference with respect to the case $\beta=0$
studied in the present paper is that
$H_{\alpha,\beta}$ can possess non-real
complex conjugate eigenvalues for $\beta\not=0$.

\section*{Acknowledgment}
%
One of the authors (D.K.) would like to thank
the foundation \emph{\v{C}esk\'y liter\'arn{\'\i} fond}
for a financial support which enabled him to participate
in the \emph{3rd International Workshop on Pseudo-Hermitian
Hamiltonians in Quantum Physics} (Istanbul, 2005).
The work has partially been supported by
the Czech Academy of Sciences and its Grant Agency
within the projects IRP AV0Z10480505 and A100480501

%
{\small

}
\end{document}